# Spatial Effects and Verdoorn Law in the Portuguese Context

Vítor João Pereira Domingues Martinho[1]

[1]Escola Superior Agrária, Instituto Politécnico de Viseu, Quinta da Alagoa,
Estrada de Nelas, Ranhados, 3500 - 606 VISEU
Centro de Estudos em Educação, Tecnologias e Saúde (CI&DETS)
**Portugal**
**e-mail:** vdmartinho@esav.ipv.pt

**ABSTRACT**

The consideration of spatial effects at a regional level is becoming increasingly frequent and the work of Anselin (1988), among others, has contributed to this. This study analyses, through cross-section estimation methods, the influence of spatial effects in productivity (product per worker) in the NUTs III economic sectors of mainland Portugal from 1995 to 1999 and from 2000 to 2005 (taking in count the availability of data), considering the Verdoorn relationship. To analyse the data, by using Moran I statistics, it is stated that productivity is subject to a positive spatial autocorrelation (productivity of each of the regions develops in a similar manner to each of the neighbouring regions), above all in services. The total of all sectors present, also, indicators of being subject to positive autocorrelation in productivity. Bearing in mind the results of estimations, it can been that the effects of spatial spillovers, spatial lags (measuring spatial autocorrelation through the spatially lagged dependent variable) and spatial error (measuring spatial autocorrelation through the spatially lagged error terms), influence the Verdoorn relationship when it is applied to the economic sectors of Portuguese regions (Martinho, 2011).

**Keywords:** Spatial Econometric; Verdoorn Law; Portuguese Regions.

## 1. INTRODUCTION

The influence of neighbouring locations (parishes, councils, districts, regions, etc) in the development of a particular area, through the effects of spatial spillovers, is increasingly considered in more recent empirical studies, a fact which has been highlighted by Anselin (2002a). Anselin (1988 and 2001) and Anselin and Bera (1998), who refer to the inclusion of spatial effects as being important from an econometric point of view. If the underlying data arises from processes which include a spatial dimension, and this is omitted, the estimators are either biased and inconsistent or inefficient depending on whether the error or the lag model is the underlying data generating process.

Following on from these studies, the development of productivity of a particular region, for example, can be influenced by the development of productivity in neighbouring regions, through external spatial factors. The existence or non-existence of these effects can be determined through a number of techniques which have been developed for spatial econometrics, where Anselin, among others, in a number of studies has made a large contribution. Paelinck (2000) has brought a number of theoretical contributions to the aggregation of models in spatial econometrics, specifically concerning the structure of parameters. Anselin (2002b) considered a group of specification tests based on the method of Maximum Likelihood to test the alternative proposed by Kelejian and Robinson (1995), related to perfecting the spatial error component. Anselin (2002c) has presented a classification of specification for models of spatial econometrics which incorporates external spatial factors. Anselin (2002d) has reconsidered a number of conceptual matters related to implementing an explicit spatial perspective in applied econometrics. Baltagi et al. (2003) has sought to present improvements in specification tests (testing whether the more correct specification of models is with the spatial lag component or the spatial error component) LM (Lagrange Multiplier), so as to make it more adaptable to spatial econometrics. Anselin et al. (1996) has proposed a simple, robust diagnostic test, based on the OLS method, for the spatial autocorrelation of errors in the presence of spatially lagged dependent variables and vice-versa, applying the modified LM test developed by Bera and Yoon (1993). This test was, also, after proposed by Florax et al. (2003).

This study seeks to test Verdoorn's Law (using product per worker as a proxy for productivity) for each of the economic sectors of regions (NUTs III) of mainland Portugal from 1995 to 1999 and from 2000 to 2005, through techniques of cross-section spatial econometrics. To do so, the rest of the study is structured as follows: in the second part some studies which have already been developed in the area of spatial econometrics, specifically concerning Verdoorn's Law, are presented; in the third part some theoretical considerations of spatial econometrics are presented; in the fourth, the models considered are explained; in the fifth the data is analysed based on techniques of spatial econometrics developed to explore spatial data; the sixth presents estimations under Verdoorn's Law, taking into account spatial effects; and in the seventh part the main conclusions obtained through this study are presented.

## 2. EMPIRICAL CONTRIBUTIONS BASED ON SPATIAL EFFECTS

There have been various studies carried out concerning Verdoorn's Law considering the possibility of there being spatial spillover effects.

Concerning Verdoorn's Law and the effects of spatial lag and spatial error, Bernat (1996), for example, tested Kaldor's three laws of growth[1] in North American regions from 1977-1990. The results obtained by Bernat clearly supported the first two of Kaldor's laws and only marginally the third. Fingleton and McCombie (1998) analysed the importance of scaled growth income, through Verdoorn's Law, with spatial lag effects in 178 regions of the European Union in the period of 1979 to 1989 and concluded that there was a strong scaled growth income. Fingleton (1999), with the purpose of presenting an alternative model between Traditional and New Geographical Economics, also constructed a model with the equation associated to Verdoorn's Law, augmented by endogenous technological progress involving diffusion by spillover effects and the effects of human capital. Fingleton applied this model (Verdoorn) to 178 regions of the European Union and concluded there was significant scaled growth income with interesting results for the coefficients of augmented variables (variable dependent on redundancy, rurality, urbanisation and diffusion of technological innovations)) in Verdoorn's equation.

Few studies have been carried out on analysing the conditional productivity convergence with spatial effects and none, at least to our knowledge, concerning productivity being dispersed by the various economic sectors. Fingleton (2001), for example, has found a spatial correlation in productivity when, using the data from 178 regions of the European Union, he introduced spillover effects in a model of endogenous growth. Abreu et al. (2004) have investigated the spatial distribution of growth rates in total factor productivity, using exploratory analyses of spatial data and other techniques of spatial econometrics. The sample consists of 73 countries and covers the period 1960-200. They found a significant spatial autocorrelation in the rates of total factor productivity, indicating that high and low values tend to concentrate in space, forming the so-called clusters. They also found strong indicators of positive spatial autocorrelation in total factor productivity, which increased throughout the period of 1960 to 2000. This result could indicate a tendency to cluster over time.

On the other hand, there is some variation in studies analysing conditional convergence of product with spatial effects. Armstrong (1995) defended that the fundamental element of the convergence hypothesis among European countries, referred to by Barro and Sala-i-Martin, was the omission of spatial autocorrelation in the analysis carried out and the bias due to the selection of European regions. Following on from this, Sandberg (2004), for example, has examined the absolute and conditional convergence hypothesis across Chinese provinces from the period 1985 to 2000 and found indications that there had been absolute convergence in the periods 1985-1990 and 1985-2000. He also found that there had been conditional convergence in the sub-period of 1990-1995, with signs of spatial dependency across adjacent provinces. Arbia et al. (2004) have studied the convergence of gross domestic product per capita among 125 regions of 10 European countries from 1985 to 1995, considering the influence of spatial effects. They concluded that the consideration of spatial dependency considerably improved the rates of convergence. Lundberg (2004) has tested the hypothesis of conditional convergence with spatial effects between 1981 and 1990 and, in contrast to previous results, found no clear evidence favouring the hypothesis of conditional convergence. On the contrary, the results foresaw conditional divergence across municipalities located in the region of Stockholm throughout the period and for municipalities outside of the Stockholm region during the 1990s.

Spatial econometric techniques have also been applied to other areas besides those previously focused on. Longhi et al. (2004), for example, have analysed the role of spatial effects in estimating the function of salaries in 327 regions of Western Germany during the period of 1990-1997. The results confirm the presence of the function of salaries, where spatial effects have a significant influence. Anselin et al. (2001) have analysed the economic importance of the use of analyses with spatial regressions in agriculture in Argentina. Kim et al. (2001) have measured the effect of the quality of air on the economy, through spatial effects, using the metropolitan area of Seoul as a case study. Messner et al. (2002) have shown how the application of recently developed techniques for spatial analysis, contributes to understanding murder amongst prisoners in the USA.

### 3. THEORETICAL CONSIDERATIONS OF SPATIAL ECONOMETRICS, BASED ON THE VERDOORN RELATIONSHIP

In 1949 Verdoorn detected that there was an important positive relationship between the growth of productivity of work and the growth of output. He defended that causality goes from output to productivity, with an elasticity of approximately 0.45 on average (in cross-section analyses), thus assuming that the productivity of work is endogenous.

Kaldor (1966 and 1967) redefined this Law and its intention of explaining the causes of the poor growth rate in the United Kingdom, contesting that there was a strong positive relationship between the growth of work productivity (p) and output (q), so that, p=f(q). Or alternatively, between the growth of employment € and the growth of output, so that, e=f(q). This is because, Kaldor, in spite of estimating Verdoorn's original relationship between the growth of productivity and the growth of industrial output (for countries of the OECD), gave preference to the relationship between the growth of work and the growth of output, to prevent spurious effects (double counting, since p=q-e). This author defends that there is a significant statistical relationship between the growth rate of employment or work productivity and the growth rate of output, with a regression coefficient belied to be between 0 and 1 ($0 \leq b \leq 1$), which could be sufficient condition for the presence of dynamic, statistically growing scale economies. The relationship between the growth of productivity of work and the growth of output is stronger in industry, given that mostly commercialised products are produced. This relationship is expected to be weaker for other sectors of the economy (services and agriculture), since services mostly produce non-transactional products (the demand for exports is the principal determining factor of economic growth, as was previously mentioned). And agriculture displays decreasing scale incomes, since it is characterised by restrictions both in terms of demand (inelastic demand) and supply (unadjusted and unpredictable supply).

---

[1] Kaldor's laws refer to the following: i) there is a strong link between the rate of growth of national product and the rate of growth of industrial product, in such a way that industry is the motor of economic growth; ii) The growth of productivity in industry and endogeny is dependent on the growth of output (Verdoorn's law); iii) There is a strong link between the growth of non-industrial product and the growth of industrial product, so that the growth of output produces externalities and induces the growth of productivity in other economic sectors.

More recently, Bernat (1996), when testing Kaldor's three laws of growth in regions of the USA from the period of 1977 to 1990, distinguished two forms of spatial autocorrelation: spatial lag and spatial error. Spatial lag is represented as follows: $y = \rho W y + X\beta + \varepsilon$, where y is the vector of endogenous variable observations, , W is the distance matrix, X is the matrix of endogenous variable observations, $\beta$ is the vector of coefficients, $\rho$ is the self-regressive spatial coefficient and $\varepsilon$ is the vector of errors. The coefficient $\rho$ is a measurement which explains how neighbouring observations affect the dependent variable. The spatial error model is expressed in the following way: $y = X\beta + \mu$, where spatial dependency is considered in the error term $\mu = \lambda W \mu + \xi$.

To resolve problems of spatial autocorrelation, Fingleton and McCombie (1998) considered a spatial variable which would capture the spillovers across regions, or, in other words, which would determine the effects on productivity in a determined region i, on productivity in other surrounding regions j, as the distance between i and j. The model considered was as follows:

$$p = b_0 + b_1 q + b_2 slp + u \text{, Verdoorn's equation with spatially} \qquad (1)$$
$$\text{lagged productivity}$$

where the variable p is productivity growth, q is the growth of output, $slp = \sum_j W_{ij} p_j$ (spatially lagged productivity variable), $W_{ij} = W_{ij}^* / \sum_j W_{ij}^*$ (matrix of distances), $W_{ij}^* = 1/d_{ij}^2$ (se $d_{ij} \leq 250 Km$), $W_{ij}^* = 0$ (se $d_{ij} > 250 Km$), $d_{ij}$ is the distance between regions i and j and u is the error term.

Fingleton (1999), has developed an alternative model, whose final specification is as follows:

$$p = \rho p_0 + b_0 + b_1 R + b_2 U + b_3 G + b_4 q + \xi \text{, Verdoorn's equation} \qquad (2)$$
$$\text{by Fingleton}$$

where p is the growth of inter-regional productivity, $p_0$ is the growth of extra-regional productivity (with the significance equal to the slp variable of the previous model), R represents rurality, U represents the level of urbanisation and G represents the diffusion of new technologies. The levels of rurality and urbanisation, symbolised by the R and U variables, are intended to indirectly represent the stock of human capital.

A potential source of errors of specification in spatial econometric models comes from spatial heterogeneity (Lundberg, 2004). There are typically two aspects related to spatial heterogeneity, structural instability and heteroskedasticity. Structural instability has to do with the fact that estimated parameters are not consistent across regions. Heteroskedasticity has to do with errors of specification which lead to non-constant variances in the error term. To prevent these types of errors of specification and to test for the existence of spatial lag and spatial error components in models, the results are generally complemented with specification tests. One of the tests is the Jarque-Bera test which tests the stability of parameters. The Breuch-Pagan and Koenker-Bassett, in turn, tests for heteroskedasticity. The second test is the most suitable when normality is rejected by the Jarque-Bera test. To find out if there are spatial lag and spatial error components in the models, two robust Lagrange Multiplier tests are used (LM$_E$ for "spatial error" and LM$_L$ for "spatial lag"). In brief, the LM$_E$ tests the null hypothesis of spatial non-correlation against the alternative of the spatial error model ("lag") and LM$_L$ tests the null hypothesis of spatial non-correlation against the alternative of the spatial lag model to be the correct specification.

According to the recommendations of Florax et al. (2003) and using the so-called strategy of classic specification, the procedure for estimating spatial effects should be carried out in six steps: 1) Estimate the initial model using the procedures using OLS; 2) Test the hypothesis of spatial non-dependency due to the omission spatially lagged variables or spatially autoregressive errors, using the robust tests LM$_E$ and LM$_L$; 3) If none of these tests has statistical significance, opt for the estimated OLS model, otherwise proceed to the next step, 4) If both tests are significant, opt for spatial lag or spatial error specifications, whose test has greater significance, otherwise go to step 5;; 5) If LM$_L$ is significant while LM$_E$ is not, use the spatial lag specification; 6) If LM$_E$ is significant while LM$_L$ is not, use the spatial error specification.

A test usually used to indicate the possibility of global spatial autocorrelation is the Moran's I test[2].

---
[2] A similar, but less well-known test is Geary's C test (Sandberg, 2004).

Moran's I statistics is defined as:

$$I = \frac{n}{S} \frac{\sum_i \sum_j w_{ij}(x_i - u)(x_j - u)}{\sum_i (x_i - u)^2}$$ , Moran's global autocorrelation test (3)

where n is the number of observations and $x_i$ and $x_j$ are the observed rates of growth in the locations i and j (with the average u). S is the constant scale given by the sum of all the distances: $S = \sum_i \sum_j w_{ij}$.

When the normalisation of weighting on the lines of the matrix for distances is carried out, which is preferable (Anselin, 1995), S equals n, since the weighting of each line added up should be equal to the unit, and the statistical test is compared with its theoretical average, I=-1/(n-1). Then I→0, when n→∞. The null hypothesis $H_0$: I=-1/(n-1) is tested against the alternative hypothesis $H_A$: I≠-1/(n-1). When $H_0$ is rejected and I>-1/(n-1) the existence of positive spatial autocorrelation can be verified. That is to say, the high levels and low levels are more spatially concentrated (clustered) than would be expected purely by chance. If $H_0$ is rejected once again, but I<-1/(n-1) this indicates negative spatial autocorrelation.

Moran's I local autocorrelation test investigates if the values coming from the global autocorrelation test are significant or not:

$$I_i = \frac{x_i}{\sum_i x_i^2} \sum_j w_{ij} x_j$$ , Moran's local autocorrelation test (4)

where the variables signify the same as already referred to by Moran's I global autocorrelation test.

## 4. VERDOORN'S MODEL WITH SPATIAL EFFECTS

Bearing in mind the previous theoretical considerations, what is presented next is the model used to analyse Verdoorn's law with spatial effects, at a regional and sector level in mainland Portugal.

As a result, to analyse Verdoorn's Law in the economic sectors in Portuguese regions the following model was used::

$$p_{it} = \rho W_{ij} p_{it} + \gamma q_{it} + \varepsilon_{it}$$ , Verdoorn's equation with spatial effects (5)

where p are the rates of growth of sector productivity across various regions, W is the matrix of distances across 28 Portuguese regions, q is the rate of growth of output, , $\gamma$ is Verdoorn's coefficient which measures economies to scale (which it is hoped of values between 0 and1), $\rho$ is the autoregressive spatial coefficient (of the spatial lag component) and $\varepsilon$ is the error term (of the spatial error component, with, $\varepsilon = \lambda W \varepsilon + \xi$). The indices i, j and t, represent the regions being studied, the neighbouring regions and the period of time respectively.

The sample for each of the economic sectors (agriculture, industry, services and the total of sectors) is referring to 28 regions (NUTs III) of mainland Portugal for the period from 1995 to 1999 and from 2000 to 2005.

## 5. DATA DESCRIPTION

The GeoDa programme was used to analyse the data, obtained from the National Statistics Institute, and to carry out the estimations used in this study. GeoDa[3] is a recent computer programme with an interactive environment that combines maps with statistical tables, using dynamic technology related to Windows (Anselin, 2003a). In general terms, functionality can be classified in six categories: 1) Manipulation of spatial data; 2) Transformation of data; 3) Manipulation of maps; 4) Construction of statistical tables; 5) Analysis of spatial autocorrelation; 6) Performing spatial regressions. All instructions for using GeoDa are presented in Anselin (2003b), with some improvements suggested in Anselin (2004).

The analysis sought to identify the existence of Verdoorn's relationship by using Scatterplot and spatial autocorrelation, the Moran Scatterplot for global spatial autocorrelation and Lisa Maps for local spatial autocorrelation. In this analysis of data and the estimations which will be carried out in part six of this study, the dependent variable of the equation used to test Verdoorn's Law is presented in average growth rates for the period considered for cross-section analysis.

### 5.1. ANALYSIS OF CROSS-SECTION DATA

The eight (Figure 1 and 2) Scatterplots presented below allow an analysis of the existence of a correlation between growth of productivity and product growth under Verdoorn's Law (equation (5)), for each of the economic sectors (agriculture, industry, services and the total of all sectors) of Portuguese NUTs III (28 regions), with average values for the period 1995 to 1999 and from 2000 to 2005.

---
[3] Available at http://geodacenter.asu.edu/

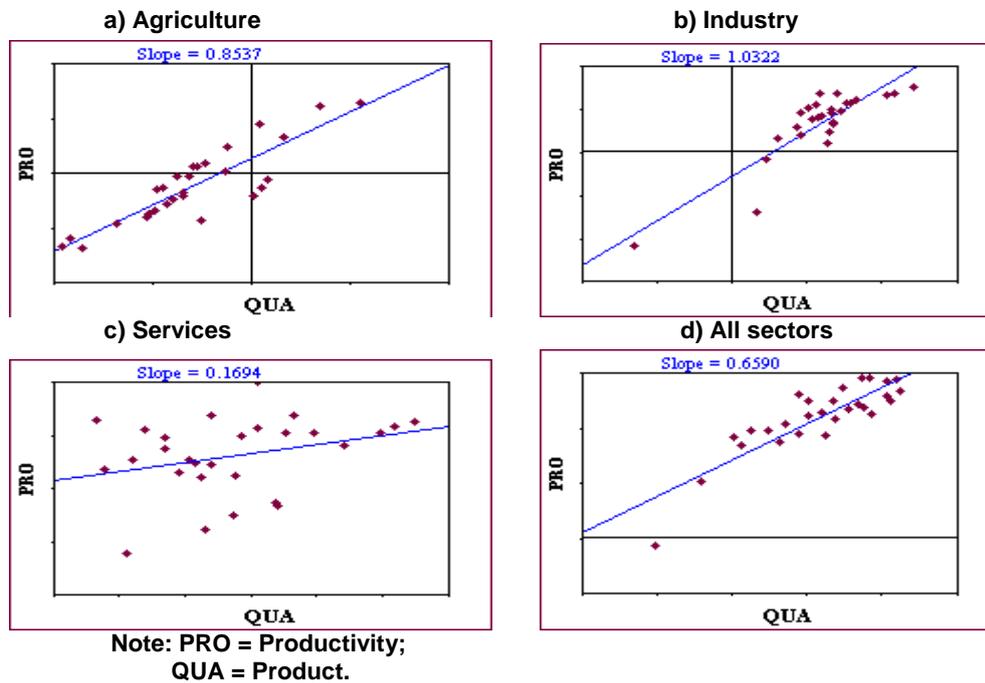

Note: PRO = Productivity;
QUA = Product.

*Figure 1:* "Scatterplots" of Verdoorn's relationship for each of the economic sector (cross-section analysis, 28 regions, 1995-1999)

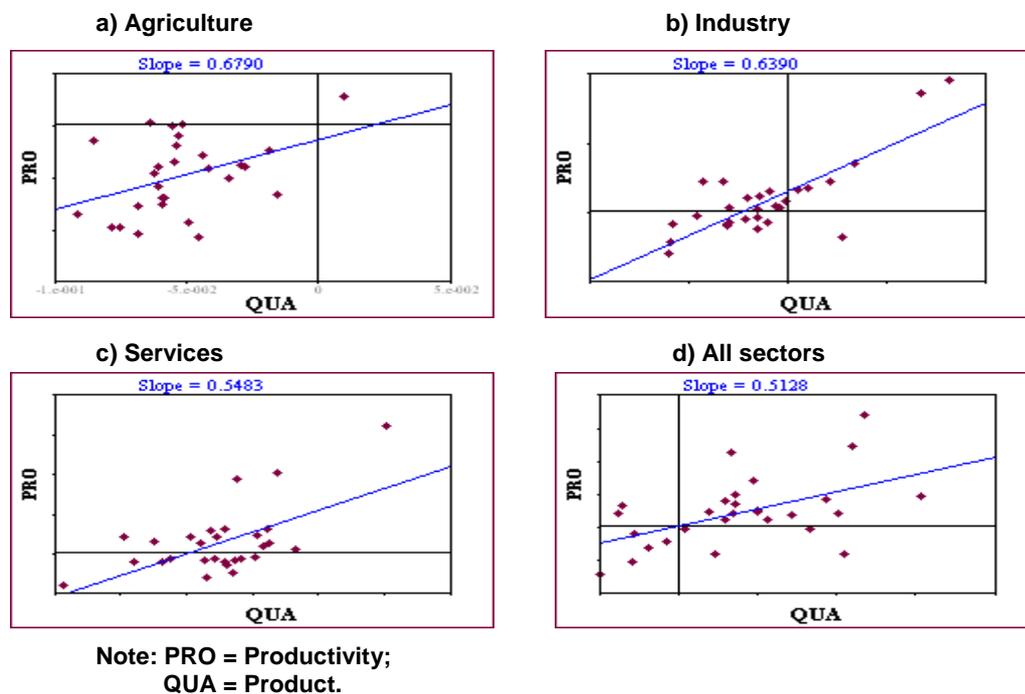

Note: PRO = Productivity;
QUA = Product.

*Figure 2:* "Scatterplots" of Verdoorn's relationship for each of the economic sector (cross-section analysis, 28 regions, 2000-2005)

To analyse the Scatterplots we confirm what is defended by Kaldor, or, in other words, Verdoorn's relationship is stronger in industry (a sign of being the sector with the greatest scaled income, although the underlying value is far too high) and weaker in other economic sectors (an indication that these sectors have less scaled income). Although agriculture is an exception here (since there is evidence of quite high scaled income, which is contrary to what was expected when considering the theory), due to the restructuring which it has undergone since Portugal joined the EEC, with the consequent decrease in population active in this sector which is reflected in increased productivity.

The eight (Figure 3 and 4) Moran Scatterplots which are presented below concerning the dependent variable (average growth rates of productivity in the period 1995 to 1999 and from 2000 to 2005), constructed by the equation of Verdoorn's Law, show Moran's I statistical values for each of the economic sectors and for the totality of sectors in the 28 NUTs in mainland Portugal. The matrix $W_{ij}$ used is the matrix of the distances between the regions up to a maximum limit of 97 Km. This distance appeared to be the most appropriate to the reality of Portuguese NUTs III, given the diverse values of Moran's I obtained after various attempts with different maximum distances. For example, for services which, as we shall see, is the sector where the Moran's I has a positive value (a sign of spatial autocorrelation), this value becomes negative when the distances are significantly higher than 97 Km, which is a sign that spatial autocorrelation is no longer present. On the other hand, the connectivity of the distance matrix is weaker for distances over 97 Km. Whatever the case, the choice of the best limiting distance to construct these matrices is always complex.

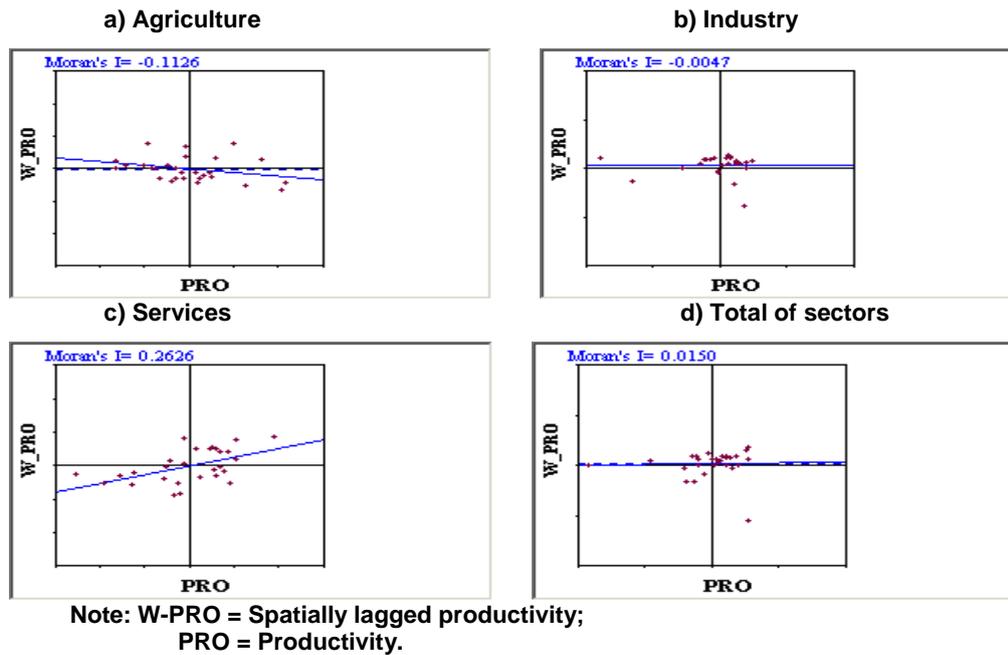

**Note: W-PRO = Spatially lagged productivity;
PRO = Productivity.**

*Figure 3:* "Moran Scatterplots" of productivity for each of the economic sectors (cross-section analysis, 28 regions, 1995-1999)

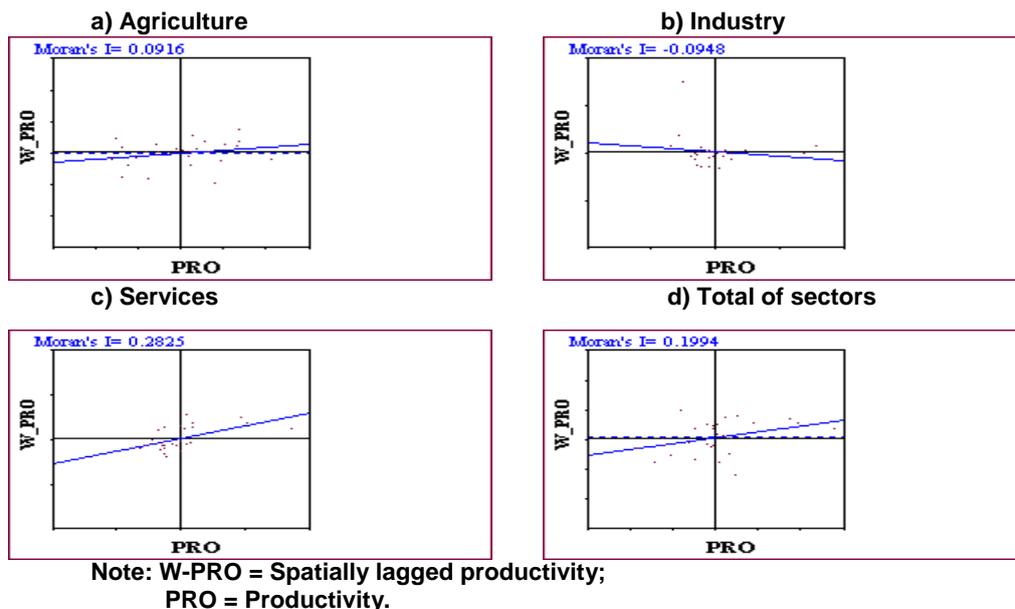

**Note: W-PRO = Spatially lagged productivity;
PRO = Productivity.**

*Figure 4:* "Moran Scatterplots" of productivity for each of the economic sectors (cross-section analysis, 28 regions, 2000-2005)

Would be good if we had more observations, but is difficult to find to a finer spatial unity. Anyway the results obtained are consistent with the Portuguese reality taking into account another works about regional growth.

An analysis of the Moran Scatterplots demonstrates that it is principally in services that a global spatial autocorrelation can be identified and that there are few indicators that this is present in the totality of sectors, since Moran's I value is positive.

Below is an analysis of the existence of local spatial autocorrelation with eight LISA Maps (Figure 5 and 6), investigated under spatial autocorrelation and its significance locally (by NUTs III). The NUTs III with "high-high" and "low-low" values, correspond to the regions with positive spatial autocorrelation and with statistical significance, or, in other words, these are cluster regions where the high values ("high-high") or low values ("low-low") of two variables (dependent variable and lagged dependent variable) are spatially correlated given the existence of spillover effects. The regions with "high-low" and "low-high" values are "outliers" with negative spatial autocorrelation. In sum, this LISA Maps find clusters for the dependent variable and lagged dependent variable.

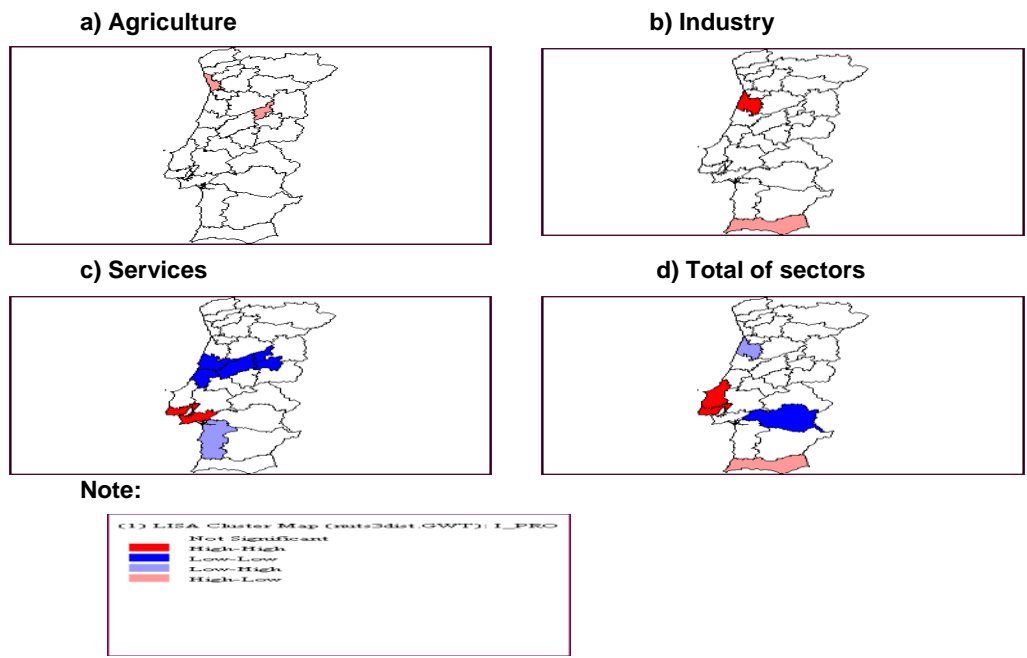

***Figure 5:*** *"LISA Cluster Map" of productivity for each of the economic sectors (cross-section analysis, 28 regions, 1995-1999)*

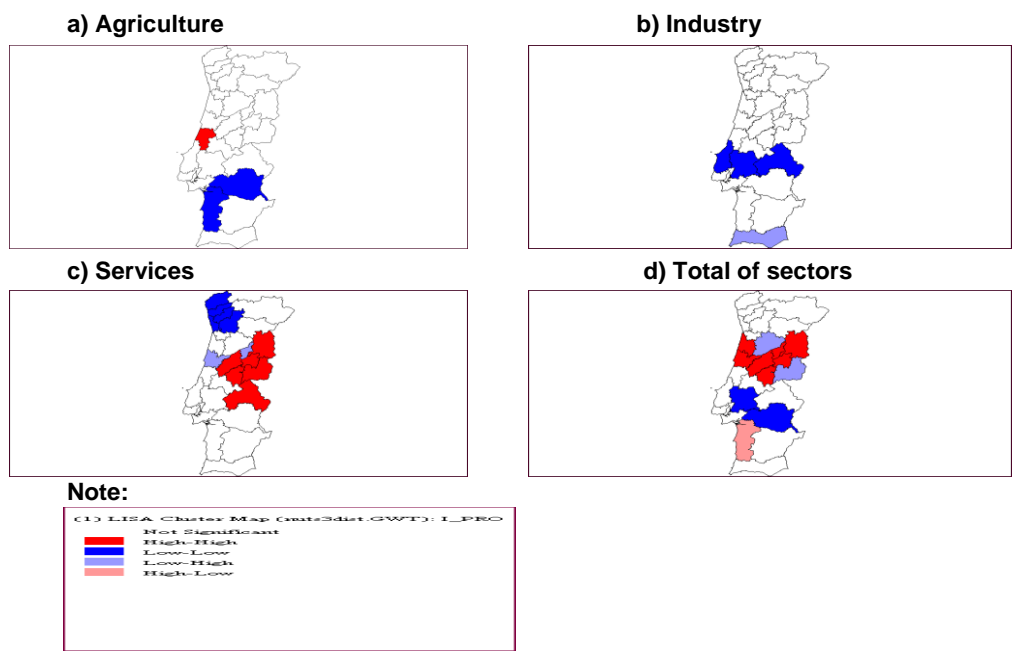

***Figure 6:*** *"LISA Cluster Map" of productivity for each of the economic sectors (cross-section analysis, 28 regions, 2000-2005)*

Upon analysing the Lisa Cluster Maps above (Figure 5), confirms what was seen with the Moran Scatterplots, or, in other words, only in the services with high values in the region around Greater Lisbon and low values in the Central region is there positive spatial autocorrelation. These figures also show some signs of positive spatial autocorrelation in all sectors, specifically with high values in the Greater Lisbon area and with low values in the Central Alentejo. Of not is the fact that industry presents signs of positive autocorrelation with high values in the Baixo Vouga in the Central region. In the second period (2000 to 2005) we can see differents situations what was expected, because the evolution of the Portuguese economy context was influenced by others factors, namely the common currency.

## 6. EMPIRICAL EVIDENCE FOR VERDOORN'S LAW, CONSIDERING THE POSSIBILITY THAT THERE ARE SPATIAL EFFECTS

The following presents empirical evidence of Verdoorn's relationship for each of the economic sectors in the Portuguese NUTs III from 1995 to 1999 and from 2000 to 2005, based on cross-section estimates. These cross- section estimates were carried out with the Least Squares (OLS) and the Maximum Likelihood (ML) methods.

### 6.1. CROSS-SECTION OF EMPIRICAL EVIDENCE

This part of the study will examine the procedures of specification by Florax e al. (2003) and will firstly examine through OLS estimates, the relevance of proceeding with estimate models with spatial lag and spatial error components with recourse to LM specification tests.

The results concerning the OLS estimates in Veroorn's equation (equation (5), without spatial variables) with spatial specification tests are presented in Tables 1 and 2. In the columns concerning the test only values of statistical relevance are presented.

**Table 1:** OLS cross-section estimates of Verdoorn's equation with spatial specification tests (1995-1999)

Equation: $p_{it} = \alpha + \beta q_{it} + \mu_{it}$

|  | Con. | Coef. | JB | BP | KB | M'I | $LM_l$ | $LMR_l$ | $LM_e$ | $LMR_e$ | $R^2$ | N.O. |
|---|---|---|---|---|---|---|---|---|---|---|---|---|
| **Agriculture** | 0.013* (3.042) | 0.854* (9.279) | 1.978 | 5.153* | 5.452* | 0.331* | 0.416 | 7.111* | 8.774* | 15.469* | 0.759 | 28 |
| **Industry** | -0.029* (-3.675) | 1.032* (9.250) | 3.380 | 2.511 | 1.532 | -0.037 | 1.122 | 2.317 | 0.109 | 1.304 | 0.758 | 28 |
| **Services** | 0.033* (3.971) | 0.169 (1.601) | 1.391 | 1.638 | 1.697 | 0.212* | 4.749* | 1.987 | 3.607* | 0.846 | 0.055 | 28 |
| **Total of sectors** | 0.002 (0.411) | 0.659* (8.874) | 1.585 | 5.174* | 4.027* | 0.030 | 0.008 | 0.087 | 0.069 | 0.149 | 0.742 | 28 |

Note: JB, Jarque-Bera test to establish parameters; BP, Breusch-Pagan test for heteroskedasticity; KB, Koenker-Bassett test for heteroskedasticity: M'I, Moran's I statistics for spatial autocorrelation; $LM_l$, LM test for spatial lag component; $LMR_l$, robust LM test for spatial lag component; $LM_e$, LM test for spatial error component; $LMR_e$, robust LM test for spatial error component; $R^2$, coefficient of adjusted determination; N.O., number of observations; *, statistically significant for 5%

**Table 2:** OLS cross-section estimates of Verdoorn's equation with spatial specification tests (2000-2005)

Equation: $p_{it} = \alpha + \beta q_{it} + \mu_{it}$

|  | Con. | Coef. | JB | BP | KB | M'I | $LM_l$ | $LMR_l$ | $LM_e$ | $LMR_e$ | $R^2$ | N.O. |
|---|---|---|---|---|---|---|---|---|---|---|---|---|
| **Agriculture** | -0.014 (-0.845) | 0.679* (2.263) | 1.201 | 0.300 | 0.505 | 0.108 | 0.771 | 0.030 | 0.940 | 0.198 | 0.132 | 28 |
| **Industry** | 0.015* (4.248) | 0.639* (6.572) | 3.238 | 2.703 | 1.393 | 0.236 | 8.742* | 4.366* | 4.444* | 0.068 | 0.610 | 28 |
| **Services** | -0.011* (-2.907) | 0.548* (3.841) | 2.728 | 9.579* | 10.452* | 0.227 | 5.976* | 1.998 | 4.102* | 0.124 | 0.338 | 28 |
| **Total of sectors** | 0.001** (0.079) | 0.513* (3.080) | 0.797 | 5.019* | 4.355* | 0.344 | 5.215* | 1.146 | 9.462* | 5.393* | 0.239 | 28 |

Note: JB, Jarque-Bera test to establish parameters; BP, Breusch-Pagan test for heteroskedasticity; KB, Koenker-Bassett test for heteroskedasticity: M'I, Moran's I statistics for spatial autocorrelation; $LM_l$, LM test for spatial lag component; $LMR_l$, robust LM test for spatial lag component; $LM_e$, LM test for spatial error component; $LMR_e$, robust LM test for spatial error component; $R^2$, coefficient of adjusted determination; N.O., number of observations; *, statistically significant for 5%

From the table 1 the existence of growing scaled income in agriculture and in the total of all sectors is confirmed. Industry shows itself to be a sector with very strong growing scaled income, since, despite Verdoorn's coefficient being highly exaggerated it is very close to unity and when the null hypothesis is tested as $\beta=1$, a t-statistic of 0.287 is obtained. As it is a highly reduced value, it is accepted that industry is subject to strong scaled income. As Kaldor predicts (1966), services are subject to constant scaled income, since, $\beta=0$. Of not is the existence of heteroskedasticity in agriculture and all sectors, given the values presented by these sectors by the BP and KB tests. As far as spatial correlation is concerned, Moran's value is only statistically significant in agriculture and services. Following the procedure of Florax et al. (2003) the equation should be estimated with the spatial error component for agriculture and with the spatial lag component for services (although in this sector none of the robust LM tests have statistical significance), with the maximum likelihood method. As expected the values obtained for the second period are significantly different.

The results for ML estimates with spatial effects for agriculture and services are presented in Tables 3 and 4.

**Table 3:** Results for ML estimates for Verdoorn's equation with spatial effects (1995-1999)

|  | **Constant** | **Coefficient** | **Coefficient$^{(S)}$** | **Breusch-Pagan** | **$R^2$** | **N.Observations** |
|---|---|---|---|---|---|---|
| **Agriculture** | 0.016* (1.961) | 0.988* (14.291) | 0.698* (4.665) | 4.246* | 0.852 | 28 |
| **Services** | 0.011 (0.945) | 0.134 (1.464) | 0.545* (2.755) | 3.050** | 0.269 | 28 |

Note: Coefficient$^{(S)}$, spatial coefficient for the spatial error model for agriculture and the spatial lag model for services; *, statistically significant to 5%; **, statistically significant to 10%.

**Table 4:** Results for ML estimates for Verdoorn's equation with spatial effects (2000-2005)

|  | **Constant** | **Coefficient** | **Coefficient$^{(S)}$** | **Breusch-Pagan** | **$R^2$** | **N.Observations** |
|---|---|---|---|---|---|---|
| **Industry** | 0.018* (5.535) | 0.682* (8.217) | -0.427* (-2.272) | 4.103* | 0.714 | 28 |
| **Services** | -0.011* (-3.308) | 0.478* (3.895) | 0.533* (2.834) | 13.186* | 0.501 | 28 |
| **All the sectors** | -0.002 (-0.379) | 0.609* (4.328) | 0.616* (3.453) | 2.230 | 0.479 | 28 |

Note: Coefficient$^{(S)}$, spatial lag model for industry and services and the spatial coefficient for the spatial error model for the all sectors; *, statistically significant to 5%; **, statistically significant to 10%.

For the first period, it is only in agriculture that Verdoorn's coefficient improves with the consideration of spatial effects, since it goes from 0.854 to 0.988. In the second period improve the Verdoorn´s coefficient of the industry and of the total sectors. Anyway, considering the presence of the spatial multiplier term in the spatial lag model the coefficients of the lag model are not directly comparable to estimates for the error model (LeSage et al., 2009 and Elhorts, 2010).

### 7. CONCLUSIONS

This study has sought to test Verdoorn's Law for each of the economic sectors (agriculture, industry, services and the totality of services) across the 28 regions (NUTs III) of mainland Portugal in the period of 1995 to 1999 and from 2000 to 2005, with spillover, spatial lag and spatial error effects. To do so, data analysis and cross-section estimates (with average temporal values) have been carried out with different estimation methods, or, in other words, OLS (least squares method) and non-linear ML (maximum likelihood method). The consideration of these two estimation methods has the objective of following the specification procedures indicated by Florax et al. (2003) who suggest that models are first tested with the OLS method, to test which is the better specification (spatial lag or spatial error) and then the spatial lag or spatial error is estimated with the GMM or ML method.

Considering the analysis of the cross-section data previously carried out, it can be seen, for the first period, that productivity (product per worker) is subject to positive spatial autocorrelation in services (with high values in the Lisbon region and low values in the Central region) and in all sectors (with high values in the Lisbon region and low values in the Central Alentejo) and also in industry (although this sector has little significance, since high values are only found in the NUT III Baixo Vouga of the Central Region). Therefore, the Lisbon region clearly has a great influence in the development of the economy with services. On the other hand, what Kaldor defended is confirmed or, in other words Verdoorn's relationship is stronger in industry, since this is a sector where growing scaled income is most expressive.

As far as cross-section estimates are concerned, it can be seen, for the first period also, that sector by sector the growing scaled income is much stronger in industry and weaker or non-existent in the other sectors, just as proposed by Kaldor. With reference to spatial autocorrelation, Moran's I value is only statistically significant in agriculture and services. Following the procedures of Florax et al. (2003) the equation is estimated with the spatial error component for agriculture and the spatial lag component for services, it can be seen that it is only in agriculture that Verdoorn's coefficient improves with the consideration of spatial effects.

For the second period the data and the results are different, what is waited, because the context in Portugal is distinct and in our point of view the indicators are better. In the first period, industry is one of the sectors with less spatial spillover effects in mainland Portugal and which has the greatest growing scaled income, because this we could conclude that the development of the national economy does not have a very favourable internal outlook with these results. So, it would be advisable to favour economic policies seeking to modernise industrial structures in Portugal, so that industry can benefit from spillover effects, as seen in services, what happened in the second period.